\documentclass[reqno]{amsart} 
\usepackage{enumerate}
\newtheorem{Theorem}{Theorem}
\newtheorem{Proposition}{Proposition}
\newtheorem{Lemma}{Lemma}                                                           
\newtheorem{Definition}{Definition}

\numberwithin{equation}{section}
 
\newcommand{\vo}{(1 + g_{rs} v^r v^s)}

\title{Asymptotic behaviour of the Einstein-Vlasov system with a positive cosmological constant}
\author{Hayoung Lee}
\address{Max-Planck-institut f\"ur Gravitationsphysik, 
Am M\"uhlenberg 1, Golm, 14476, Germany}
\email{hayoung@aei.mpg.de} 
\begin{document}

\begin{abstract}
We study locally spatially homogeneous solutions of the Einstein-Vlasov system with a positive cosmological constant. 
First the global existence of solutions of this system and 
the casual geodesic completeness are shown. Then the asymptotic behaviour of solutions in
the future time is investigated in various aspects.
\end{abstract}

\maketitle
\markboth{\sc Hayoung Lee}{\sc Asymptotic Behaviour of the Einstein-Vlasov system}

%%####################################################################################
\section{Introduction}
The Vlasov equation in general describes a collection of collisionless particles.
Each particle is driven by self-induced fields which are generated by all particles
together. In gravitational physics, which is the subject we are considering here,
the fields are described by the Einstein equations. 
In other words, the Vlasov equation is coupled to the Einstein equations
by using the energy-momentum tensor. One application of the Vlasov
equation to this self-gravitating system is cosmology.
And the particles are in this case galaxies or even clusters of galaxies.

We study the dynamics of expanding cosmological models in the presence of a positive cosmological constant.
The simplest cosmological models are those which are spatially homogeneous.
Spatially homogeneous spacetimes can be classified into two types. 
The conventional terminology is that there are Bianchi models and
the Kantowski-Sachs models. 
The models with a three-dimensional group of isometries $G_3$ acting simply
transitively on spacelike hypersurfaces are Bianchi models. There are nine types I-IX, 
depending upon the classification of the structure constants of the Lie algebra of $G_3$.
Those admitting a group of isometries $G_4$ which acts on spacelike hypersurfaces but no subgroup
$G_3$ which acts transitively on the hypersurface are Kantowski-Sachs models.
In fact, $G_3$ subgroup acts multiply transitively on two-dimensional spherically symmetric surfaces.
See \cite{WainEllis, Wald2} for
general information on this classification.
 
At this point it is necessary to discuss which solutions of the Einstein-Vlasov system
should be considered as cosmological solutions. If we take as a cosmological spacetime
one which admits a compact Cauchy hypersurface, the Bianchi types which
can occur for a spatially homogeneous cosmological model are only type I and IX and also
Kantowski-Sachs models.
Because of the existence of {\em locally} spatially homogeneous cosmologies, 
we take a larger class of spacetimes possessing a compact Cauchy hypersurface
so that this allows a much bigger class of Bianchi types to be included.
Since the Cauchy problem for the Einstein-Vlasov system is well-posed, 
it is enough to define the class of initial data.
Here is the definition.
\begin{Definition}
{\rm Let $\stackrel{\rm _o}{g}_{ij}$, $\stackrel{\rm _o}{k}_{ij}$ and $\stackrel{\rm _o}{f}$ be initial data for a Riemannian metric, 
a second fundamental form, and the Vlasov equation, respectively, 
on a three-dimensional manifold $M$.
Then this initial data set $(\stackrel{\rm _o}{g}_{ij}, \stackrel{\rm _o}{k}_{ij}, \stackrel{\rm _o}{f})$ 
for the Einstein-Vlasov system is called {\em locally spatially homogeneous} 
if the naturally associated data set on the universal covering $\widetilde{M}$ is
homogeneous, i.e. invariant under a transitive group action.}
\end{Definition}
So the spacetimes considered here will be Cauchy developments
of locally homogeneous initial data sets on some manifolds.
Singer's main theorem in \cite{Sing} says
\begin{quote}
\it If $M$ is a complete, simply connected Riemannian manifold which is locally homogeneous,
then $M$ is a homogeneous Riemannian manifold.
\end{quote}
So one can see that a complete Riemannian manifold is
locally homogeneous if and only if the universal cover is homogeneous.
For Bianchi models the universal covering space can be identified with a Lie group $G$.
So the natural choice for $G$ in this case is a simply connected three-dimensional Lie group.
(For a detailed discussion on this we refer to \cite{Ren1, Ren2}).

Without a cosmological constant, it is known that solutions of Bianchi type IX 
and Kantowski-Sachs models cannot
expand forever, while the other Bianchi types force the volume to be monotone. 
In fact, if the Bianchi type is IX  or the model is Kantowski-Sachs,
then it has curvature singularities after finite proper time
both in the past and in the future directions. All other Bianchi types which
are expanding at some time have a curvature singularity at a finite time in the past and 
all these are future geodesically complete (see \cite{Ren1, Ren2}).

The detailed picture of the asymptotics of the spatially homogeneous cosmological models without
a cosmological constant has not yet been achieved in general. 
However in \cite{RenTod} for the Einstein-Vlasov system with massless particles the LRS 
(locally rotationally symmetric) reflection-symmetric solutions of Bianchi types I, II, III, $\rm VII_0$ 
and Kantowski-Sachs type have been analysed as far as to give a full description of 
their general behaviour near the singularity and in a phase of unlimited expansion.
Massive particles for LRS reflection-symmetric solutions of Bianchi types I, II and III are handled in \cite{RenUgg}.
Also in \cite{Ren5} LRS models of Bianchi type III are analysed, which extends and completes
the results of \cite{RenTod} and \cite{RenUgg}.
In the case where the matter model considered is a perfect fluid, much more is known.

In the presence of a positive cosmological constant, in \cite{TchRen} 
the existence of {\em inhomogeneous} solutions of the Einstein-Vlasov system is proved
and future asymptotic behaviours are investigated, in the cases of plane and hyperbolic symmetries.
The spatially homogeneous cases of the results of the plane and hyperbolic symmetries in \cite{TchRen} 
correspond to LRS models of Bianchi type I and LRS of Bianchi III respectively.
Surveys of results on the Einstein-Vlasov system in general 
and other cosmological models can be found in \cite{Andre, Ren6, Ren3, Ren4, Ren7, Ren8}.

When a positive cosmological constant is present, Bianchi IX and Kantowski-Sachs models have complicated features.
These have both expanding and recollapsing phases.
One can find some results on Kantowski-Sachs models in \cite{M}. In \cite{GE} Kantowski-Sachs model with perfect
fluid matter is analysed.
In \cite{B9} Bianchi IX with dust model has been studied. It has been found that
there are chaotic behaviours between expanding and recollapsing phases. 
In this paper, in discussing expanding cosmological models, we restrict ourselves to Bianchi types except IX.

At late times Bianchi models except IX with a cosmological constant must expand exponentially. 
This kind of exponential expansion is called {\em inflation}. 
The inflationary models of the universe involve a very rapid expansion 
close to the big bang. Also the fact that the expansion of our universe seems to be accelerating 
follows from observations of supernovae of type Ia. 

The content of the rest of this paper is the following. In Section \ref{exist.geodesic},
we discuss the global existence of solutions for the Einstein-Vlasov system and 
the causal geodesic completeness of the spacetime towards the future.
In Section \ref{asymp}, we investigate the asymptotic behaviour of solutions at late times in various aspects.
As we will see later on, in the case where a positive cosmological constant is present,
the situations of the locally spatially homogeneous cosmological models have a unified picture in their 
expanding dynamics.
We observe the future asymptotic behaviours of the mean curvature, the metric,
the momenta of particles along the characteristic curves as well as the generalized Kasner
exponents and the deceleration parameter. 
Also we analyse the energy-momentum tensor in an orthonormal frame on the hypersurfaces.

Here is the formulation of the Einstein-Vlasov system with a positive cosmological constant.
Let $G$ be a simply connected 
three-dimensional Lie group and $\{e_i\}$, a left invariant frame 
and $\{e^i\}$, the dual coframe.
Consider the spacetime as a manifold $G \times I$, where $I$ is an open interval and 
the spacetime metric of our model has the form
\begin{equation}\label{metric}
ds^2 = -dt^2 + g_{ij}(t) e^{i} \otimes e^{j}.
\end{equation}
The initial value problem for the Einstein-Vlasov system is investigated in the case of this special form 
of the metric and the distribution function $f$ depends only on $t$ and $v^i$, 
where $v^i$ are spatial components of the momentum in the frame ${e_i}$. 
Initial data will be given on the hypersurface $G \times \{0\}$.

Now the constraints are
\begin{gather}
R-(k_{ij}k^{ij}) + (k_{ij}g^{ij})^2 = 16 \pi T_{00} + 2\Lambda, \label{cons1}\\
\nabla^i k_{ij} = - 8 \pi T_{0j}. \label{cons2}
\end{gather}
The evolution equations are
\begin{align}
\partial_t g_{ij} & = -2 k_{ij}, \label{evol1}\\
\partial_t k_{ij} & = R_{ij} + (k_{lm}g^{lm}) k_{ij} - 2 k_{il} k^l_j - 8 \pi T_{ij} \label{evol2}\\
		  & \qquad\qquad  - 4\pi T_{00} g_{ij} + 4 \pi (T_{lm}g^{lm}) g_{ij} - \Lambda g_{ij}. \notag
\end{align}
These equations are written using frame components. Here $k_{ij}$ is
the second fundamental form, $R$ is the Ricci scalar curvature and $R_{ij}$ is the Ricci
tensor of the three-dimensional metric. And $\Lambda$ is a positive cosmological constant.

Here are the components of the energy-momentum tensor ;
\begin{align}
T_{00} (t) & = \int f(t, v) \vo^{1/2} (\det g)^{1/2} \,dv\label{tensor1}\\
T_{0i} (t) & = \int f(t, v) v_i (\det g)^{1/2} \,dv\label{tensor2}\\
T_{ij} (t) & = \int f(t, v) v_i v_j \vo^{-1/2} (\det g)^{1/2} \,dv.\label{tensor3}
\end{align}
Here $v:=(v^1, v^2, v^3)$ and $dv :=dv^1\,dv^2\,dv^3$.

The Vlasov equation is
\begin{equation} \label{vlasov}
\partial_t f + 
  \big\{ 2 k^i_j v^j - \vo^{-1/2}\gamma^i_{mn} v^m v^n \big\} \partial_{v^i}f = 0
\end{equation}
Here the Ricci rotation coefficients $\gamma^i_{mn}$ are defined as
$$
\gamma^i_{mn} = \frac{1}{2} g^{ik} (-C^l_{nk} g_{ml} + C^l_{km} g_{nl} + C^l_{mn} g_{kl})
$$
where $C^i_{jk}$ are the structure constants of the Lie algebra of $G$.

In general the evolution equations are partial differential equations. However since the spacetime is locally
spatially homogeneous in this paper, 
this gives a big simplification. I.e., $\partial_t$ is actually $d/dt$ in the evolution
equations. Yet the Vlasov equation remains as a partial differential equation.
We abuse the notation in this way to follow the notation in the literature.

%%###############################################################################
%% section : the global existence of solutions and the geodesic completeness
%%###############################################################################
\section{The global existence of solutions and\\
 the future geodesic completeness}\label{exist.geodesic}

In the absence of a cosmological constant the existence of solutions 
for the Einstein-Vlasov system has been proved in \cite{Ren1} by A.~D.~Rendall under some conditions.
First under the following assumptions on the initial data :
\begin{enumerate}[(1)]
\item $g_{ij}(0)$, $k_{ij}(0)$ and $f(0, v)$ are initial data
for equations (\ref{evol1}), (\ref{evol2}) and (\ref{vlasov}) which have Bianchi symmetry and
satisfy the constraints (\ref{cons1}) and (\ref{cons2}),
\item $f(0, v)$ is a $C^1$ function of compact support, 
\end{enumerate}
the local existence of solutions has been proved.
The argument to achieve the global existence is that,
if $|g|$, $(\det g)^{-1}$, $|k|$, $\|f\|_\infty$ and the diameter of the support of $f$ are bounded on
the finite time interval, $[0, T)$ where the solutions exist locally, then $T=\infty$. 
Here $g$ and $k$ are matrices with components $g_{ij}$ and $k_{ij}$ respectively. 
And $|g|$ denotes the maximum modulus of any components $g_{ij}$ for all $t$, with the similar definition for $|k|$. 
Another criterion for the global existence, which is a stronger version,
is the following. If the mean curvature is bounded on $[0, T)$,
then $|g|$, $(\det g)^{-1}$, $|k|$, $\|f\|_\infty$ and the diameter of the support $f$ are also bounded in $[0, T)$.
In another words, the boundedness on the trace $k$ implies $T=\infty$, i.e.,
solutions of the system exist globally.
Here the existence of solutions means the existence of solutions and the uniqueness.

In the case of the presence of the cosmological constant, to obtain the local
existence of solutions, these results from \cite{Ren1}
can be directly applied with a minor change. So it is enough to obtain
the boundedness of the mean curvature to achieve the existence of the global solutions.

Here is the statement of the global existence of solutions for the Einstein-Vlasov system with a cosmological constant.

%%############################################
% thm : global existence of solution
%%############################################
\begin{Theorem}\label{thm:exist}
Let $(g_{ij}(0), k_{ij}(0), f(0, v))$ be an initial data set for the evolution equations (\ref{evol1}),
(\ref{evol2}) and the Vlasov equation (\ref{vlasov}) which has Bianchi symmetry and 
satisfies the constraints (\ref{cons1}) and (\ref{cons2}). 
Also let $f(0, v)$ be a nonnegative $C^1$ function with compact support.
Then there exists a unique $C^1$ solution $(g_{ij}, k_{ij}, f)$ of the Einstein-Vlasov
system, for all time $t$.
\end{Theorem}
%%################################################################
{\sc Proof.} (\ref{evol1})-(\ref{vlasov}) imply a homogeneous first
order ordinary differential system for (\ref{cons1}) and (\ref{cons2}).
To be more precise, define
\begin{align*}
A &:= R-(k_{ij}k^{ij}) + (k_{ij}g^{ij})^2 - 16 \pi T_{00} - 2\Lambda, \\
A_i &:= \nabla^i k_{ij} + 8 \pi T_{0j}.
\end{align*}
Then after a lengthy calculation we obtain
\begin{gather*}
\partial_t A = 2(k_{ij}g^{ij})A -2 \gamma^l_{ij}g^{ij}A_l,\\
\partial_t A_i = (k_{lm}g^{lm})A_i + 2 k^l_i A_l.
\end{gather*}
This means that if the initial data satisfy the constraints (\ref{cons1}) and (\ref{cons2})
then so does the solution of the evolution equations (\ref{evol1}) and (\ref{evol2})
with energy-momentum tensors (\ref{tensor1})-(\ref{tensor3}) and the Vlasov equation
(\ref{vlasov}).
As we have mentioned before, by the same argument in \cite{Ren1} (see pp. 86-89), 
the solution is uniquely determined by the initial data on $[0, T)$, for some time $T$.
So we need to check the boundedness of the mean curvature to achieve the global solution.
This follows from (\ref{kijgij}) in Proposition \ref{kg+sigma} which we state and prove
in the next section where it is mainly used.\hfill$\Box$

The next result asserts the geodesic completeness of locally spatially homogeneous spacetimes for
the Einstein-Vlasov system with a positive cosmological constant.

%%########################################
%% thm : geodesic completeness
%%########################################
\begin{Theorem}\label{thm.geodesic}
Suppose the hypotheses of Theorem \ref{thm:exist} hold. Then the spacetime is future complete.  
\end{Theorem}
%%#########################################
We postpone the proof of this theorem to the next section.

%%###########################################################
%% section : asymptotic behaviours of solutions
%%###########################################################
\section{The future asymptotic behaviour of solutions}\label{asymp}

Bianchi models except type IX with a cosmological constant correspond to exponential
inflation. Also the spacetime becomes isotropic at late times.
In other words, all locally spatially homogeneous solutions of the Einstein-Vlasov system with a positive
cosmological constant are approximated by the de Sitter solution as time goes to infinity. 
The idea that the de Sitter solution should act the attractor
for solutions of the Einstein equations with a cosmological constant 
is called {\em the cosmic no hair conjecture}. Wald's result (\cite{Wald1}) can be
considered as a proof of this conjecture in the homogeneous case.
In \cite{Wald1} the form of the energy-momentum tensor is not specified
in detail. It satisfies only the dominant and strong energy conditions
given respectively by
\begin{enumerate}[(1)]
\item $T_{\alpha\beta} v^\alpha w^\beta \geq 0$ where $v^\alpha$ and $w^\beta$ are
any two future pointing timelike vectors,
\item $R_{\alpha\beta} v^\alpha v^\beta \geq 0$ for any timelike vector $v^\alpha$.
\end{enumerate}
Here Greek indices take the values 0, 1, 2, 3.

In this section, we investigate the asymptotic behaviour of solutions of the Einstein-Vlasov system in the future
time. The basic ideas of the following proposition are from \cite{Wald1}. 
Here we carry out the computation carefully
so that the error terms are explicitly determined for the future reference. 
In general throughout the paper,
a constant $C$ is positive and may change line by line.

%%######################################
%% 		k_ij g^ij ,  sigma_ij
%%######################################
\begin{Proposition}\label{kg+sigma}
Let $\sigma_{ij}$ be the trace free part of the second fundamental form $k_{ij}$ such that
\begin{equation}  	 	 	   			 			   \label{kij}
k_{ij} = \frac{1}{3}(k_{lm}g^{lm}) g_{ij} + \sigma_{ij}.
\end{equation}
Then we have
\begin{gather}
(k_{ij}g^{ij}) = -(3\Lambda)^{1/2} + \mathcal{O}(e^{-2\gamma t}) \label{kijgij}\\
(\sigma_{ij}\sigma^{ij}) = \mathcal{O}(e^{-2\gamma t}) \label{sigmaijij}
\end{gather}
where $\gamma^2 = \Lambda/3$.
\end{Proposition}
%%#######################################
{\sc Proof.} 
From evolution equations (\ref{evol1}) and (\ref{evol2}), we have
\begin{equation} \label{evol3}
\partial_t(k_{ij}g^{ij}) 
	= R + (k_{ij}g^{ij})^2 + 4 \pi (T_{ij} g^{ij}) - 12\pi T_{00} - 3\Lambda.
\end{equation}
Using the constraint (\ref{cons1}), (\ref{evol3}) becomes
\begin{equation}\label{evol4}
\partial_t (k_{ij}g^{ij}) = (k_{ij}k^{ij}) + 4\pi (T_{ij}g^{ij}) + 4\pi T_{00} -\Lambda.
\end{equation}
By the dominant and strong energy conditions and 
using (\ref{kij}), we obtain from (\ref{evol4}) the following :
\begin{equation} \label{evol5}
\partial_t(k_{ij}g^{ij}) \geq (k_{ij}k^{ij}) - \Lambda 
						 \geq \frac{1}{3}(k_{ij}g^{ij})^2 - \Lambda.
\end{equation}
Also we rewrite the constraint (\ref{cons1}) using (\ref{kij}) as
\begin{equation}\label{cons11}
\frac{1}{3}(k_{ij}g^{ij})^2 = -\frac{1}{2}R + \frac{1}{2}(\sigma_{ij} \sigma^{ij})
				   + 8 \pi T_{00} + \Lambda.
\end{equation}
In \cite{Wald1}, Wald has proved that in all Bianchi models except type IX, 
the Ricci scalar curvature is zero or negative.
So from (\ref{cons11}), we can obtain that the last quantity in (\ref{evol5}) is nonnegative, i.e.
\begin{equation}
\partial_t(k_{ij}g^{ij}) \geq \frac{1}{3}(k_{ij}g^{ij})^2 - \Lambda \geq 0. \label{ineq}
\end{equation}
From the last inequality,
one can see that if the universe is initially expanding, i.e., the mean curvature
is nonzero initially which will be the case here, then the mean curvature never becomes zero and this means
that the universe expands forever. 
Now the first case for the solutions from (\ref{ineq}) is 
by taking the equal sign in the last inequality :
\begin{equation} \label{kgforT}
(k_{ij}g^{ij}) = -(3\Lambda)^{1/2}
\end{equation}
for all time $t$.
We choose the negative sign as the convention throughout the paper.
The other case is by solving the following ordinary differential inequality :
\begin{equation*}
\frac{\partial_t(k_{ij}g^{ij})}{(k_{ij}g^{ij})^2-3\Lambda} \leq \frac{1}{3}.
\end{equation*}
I.e., integrating both side of the following
\begin{equation*}
\partial_t \left(\ln\frac{(k_{ij}g^{ij})-(3 \Lambda)^{1/2}}{(k_{ij}g^{ij})+(3 \Lambda)^{1/2}} \right) \leq 2 \gamma.
\end{equation*}
we obtain
\begin{equation*}
0 > (k_{ij}g^{ij}) + (3\Lambda)^{1/2} \geq - C e^{-2\gamma t}
\end{equation*}
where $C$ is a positive constant.
Therefore combining with (\ref{kgforT}), we get
\begin{equation}\label{kgforO}
(k_{ij}g^{ij}) = -(3\Lambda)^{1/2} + \mathcal{O} (e^{-2\gamma t}).
\end{equation}
Here while we use the notation $\mathcal{O}(\cdot)$, we lose a piece of information from (\ref{ineq}) that
the mean curvature is non-decreasing. So we want to point out that in (\ref{kgforO})
the error $\mathcal{O} (e^{-2\gamma t})$ is possibly zero or negative error.

Now to prove (\ref{sigmaijij}), we use (\ref{cons11}) once more. Then we get
\begin{equation*}
\frac{1}{2}(\sigma_{ij} \sigma^{ij}) \leq \frac{1}{3}(k_{ij}g^{ij})^2 - \Lambda.
\end{equation*}
Using (\ref{kijgij}), one can see that
\begin{equation*}
(\sigma_{ij}\sigma^{ij}) = \mathcal{O} (e^{-2\gamma t}).
\end{equation*}
\hfill$\Box$

To identify $\sigma_{ij}$ from (\ref{sigmaijij}) which is necessary to investigate the asymptotic
behaviour of $g_{ij}$,
let us use some properties of linear algebra which can be found in \cite{Ren1}. 
Let $A_1$ and $A_2$ be $n \times n$ symmetric matrices with $A_1$ positive definite. Also let $A$
be a $n \times n$ matrix. Define {\em a norm} of a matrix by
\begin{equation*}
\|A\| := \sup \{\|Ax\|/\|x\| : x \neq 0,\, x \in \mathbb{R}^n\}.
\end{equation*}
Also define {\em a relative norm} by 
\begin{equation*}
\| A_2 \|_{A_1} := \sup \{ \|A_2 x \| / \|A_1 x \| : x \neq 0 , \, x \in \mathbb{R}^n \}.
\end{equation*}
Then from these definitions, one can see that
\begin{equation}\label{norm}
\|A_2\| \leq \|A_2\|_{A_1}\|A_1\|
\end{equation}
and also
\begin{equation}\label{relnorm}
\|A_2\|_{A_1} \leq \big( \text{tr}(A_1^{-1} A_2 A_1^{-1} A_2) \big)^{1/2}.
\end{equation}

Here is a lemma which gives a relation between $\sigma_{ij}$ and $g_{ij}$.

%%##################################
%% lemma for \sigma_{ij}
%%##################################
\begin{Lemma} \label{lem.sigmat}
Let $\|g(t)\|$, $\|k(t)\|$ and $\|\sigma(t)\|$ denote the norms of the matrices with entries 
$g_{ij}$, $k_{ij}$ and $\sigma_{ij}$, respectively. Then
\begin{equation*}
\|\sigma(t)\| \leq C e^{-\gamma t} \| g(t) \|
\end{equation*}
where $C$ is a constant.
\end{Lemma}

{\sc Proof.} By (\ref{relnorm}) and (\ref{sigmaijij}) we have
\begin{equation*}
\|\sigma(t)\|_{g(t)} \leq (\sigma_{ij} \sigma^{ij})^{1/2} \leq C e^{- \gamma t}.
\end{equation*}
So the lemma follows by (\ref{norm}). \hfill $\Box$

The next lemma gives some information on $g_{ij}$ and $g^{ij}$. It does not say the full information
about the asymptotic behaviours of $g_{ij}$ and $g^{ij}$. 
It is a step in an argument which leads to stronger estimates on these quantities.

%%###################################
%% bounds on g_ij g^ij
%%###################################
\begin{Lemma} \label{egij}
\begin{align}
|e^{-2 \gamma t} g_{ij}| & \leq C,\\
|e^{2 \gamma t} g^{ij}| & \leq C, \label{2ndgij}
\end{align}
for all $ t \geq 0$. Here $C$ is a constant.
\end{Lemma}

{\sc Proof.} 
Let $\bar{g}_{ij} := e^{- 2 \gamma t} g_{ij}$. Then using (\ref{kij}) and (\ref{kijgij}),
we get
\begin{align}
\partial_t \bar{g}_{ij} 
&= -2\gamma \bar{g}_{ij} - \frac{2}{3}(k_{lm} g^{lm}) \bar{g}_{ij} - 2 e^{-2 \gamma t} \sigma_{ij} \notag\\
&= \mathcal{O}(e^{-2 \gamma t}) \bar{g}_{ij} - 2 e^{-2 \gamma t} \sigma_{ij}. \label{bargij}		   				
\end{align}
Now let us use the norms again. Let $\|\bar{g}(t)\|$ be a norm of the matrix with entries 
$\bar{g}_{ij}$. Then using (\ref{norm}) we get
\begin{align*}
\|\bar{g}(t)\| &\leq \|\bar{g}(0)\| + \int^t_0  C e^{-2 \gamma s} 
			   		 \big(\|\bar{g}(s)\| + \|\sigma(s)\|\big)\, ds\\
			   &\leq \|\bar{g}(0)\| + \int^t_0  C e^{-2 \gamma s} 
			   		 \big(\|\bar{g}(s)\| + \|\sigma(s)\|_{\bar{g}(s)}\|\bar{g}(s)\|\big)\, ds.
\end{align*}
Note that
\begin{equation*}
\|\sigma(s)\|_{\bar{g}(s)} \leq e^{2 \gamma t} (\sigma_{ij} \sigma^{ij})^{1/2}.
\end{equation*}
So with (\ref{sigmaijij}) we have
\begin{equation*}
\|\bar{g}(t)\| \leq \|\bar{g}(0)\| + \int^t_0  C e^{- \gamma s} \|\bar{g}(s)\| \, ds.
\end{equation*}
By Gronwall's inequality, we obtain
\begin{equation*}
\|\bar{g}(t)\| \leq \|\bar{g}(0)\| \text{exp}\left[ \int^t_0 C e^{- \gamma s} \, ds \right] \leq C
\end{equation*}
where $C$ is a constant. 
Therefore $\bar{g}_{ij}$ is bounded by a constant for all $t \geq 0$. (\ref{2ndgij}) can be proved by the same argument.
So we omit it. \hfill $\Box$

Combining Lemmas \ref{lem.sigmat} and \ref{egij}, we obtain the following proposition.
%%##################################
%% simga
%%##################################
\begin{Proposition}\label{sigmaijt}
\begin{equation*}
\sigma_{ij}= \mathcal{O} (e^{\gamma t})
\end{equation*}
for large $t$.
\end{Proposition}

The Bianchi models except type IX with a cosmological
constant are inflationary cosmological models.
I.e., the metric $g_{ij}$ increases exponentially which we have already seen in
Lemma \ref{egij}. The next proposition gives us detailed information on the error term of $g_{ij}$
which allows us to construct $g^{ij}$.
I.e. in order to find $g^{ij}$, determining the form of the error term in $g_{ij}$ is crucial.
The first main point of this proposition is to find $g_{ij}$ with the error term, 
using Propositions \ref{kg+sigma} and \ref{sigmaijt}. And then we want to obtain
$g^{ij}$ explicitly which will play an important role in the rest of the paper.

%%###############################
%%      g_{ij}, g^{ij}
%%###############################
\begin{Proposition}\label{prop_g}
\begin{align}
g_{ij}(t) &= e^{2\gamma t}(\mathcal{G}_{ij} + \mathcal{O}(e^{-\gamma t})) \label{g_ij}\\
g^{ij}(t) &= e^{-2\gamma t}(\mathcal{G}^{ij} + \mathcal{O}(e^{-\gamma t})) \label{g^ij}
\end{align}
for large $t$.
Here $\mathcal{G}_{ij}$ and $\mathcal{G}^{ij}$ are independent of t.
\end{Proposition}
%%################################
{\sc Proof.}
Consider (\ref{bargij}) again with Lemma \ref{egij} and Proposition \ref{sigmaijt}. Then we have
\begin{align*}
\partial_t \bar{g}_{ij} &= \mathcal{O}(e^{-2 \gamma t}) \bar{g}_{ij} - 2 e^{-2 \gamma t} \sigma_{ij}\\
		   				&= \mathcal{O}(e^{- \gamma t}).
\end{align*}
Since $\partial_t\bar{g}_{ij}$ is decaying exponentially,
there exists a limit, say $\mathcal{G}_{ij}$ , of $\bar{g}_{ij}$ as $t$ goes to infinity.
Then this gives
\begin{equation*}
\bar{g}_{ij} = \mathcal{G}_{ij} + \mathcal{O}(e^{- \gamma t})
\end{equation*}
i.e.
\begin{equation*}
g_{ij} = e^{2 \gamma t} \big( \mathcal{G}_{ij} + \mathcal{O}(e^{- \gamma t}) \big).
\end{equation*}

Here the lower order term of $g_{ij}$ is of an exponential form, i.e. $\mathcal{O}(e^{\gamma t})$
so that it is combined with the leading order term, which makes it possible to compute $g^{ij}$ explicitly. 
So $g^{ij}$ is
\begin{align*}
g^{ij} &= e^{-2\gamma t} \big( \mathcal{G}_{ij} + \mathcal{O}(e^{- \gamma t}) \big)^{-1}\\
		  &= e^{-2\gamma t} \big(\mathcal{G}^{ij} + \mathcal{O}(e^{-\gamma t})\big)
\end{align*}
for large time $t$. Note that the last equality is due to Taylor expansion.
\hfill$\Box$

Now let us consider the generalized Kasner exponents.
Let $\lambda_i$ be the eigenvalues of $k_{ij}$ with respect to $g_{ij}$, i.e., the solutions of
$\det(k^i_j - \lambda \delta^i_j)=0$.
Define {\em the generalized Kasner exponents} by
$$p_i := \frac{\lambda_i}{\sum_l \lambda_l} = \frac{\lambda_i}{(k_{lm}g^{lm})}.$$
The name comes from the fact that in the special case of the Kasner solutions
these are the Kasner exponents. Note that while the Kasner exponents are constants,
the generalized Kasner exponents are in general functions of $t$.
The generalized Kasner exponents always satisfy the first of the two Kasner
relations, but in general do not satisfy the second,
where these two Kasner relations are
\begin{gather}
\sum_i p_i = 1, \label{kasner_rel1}\\
\sum_i (p_i)^2 = 1.
\end{gather}
The following proposition says that all generalized Kasner exponents $p_i$ go to $\frac{1}{3}$
as $t$ goes to infinity. This means that the spacetime isotropizes at late times.

%%#######################################
%%  for generalized Kasner exponents
%%#######################################
\begin{Proposition}
$$
p_i(t) = \frac{1}{3} + \mathcal{O}(e^{-\gamma t})
$$
for large $t$.
\end{Proposition}
%%#######################################
{\sc Proof.}
First note that by (\ref{kij}) $\lambda_i$ are also the solutions of
$$
\det\big( \sigma^i_j - [\lambda - \frac{1}{3} (k_{lm}g^{lm})] \delta^i_j\big)=0.
$$
So the eigenvalues of $\sigma_{ij}$ with respect to $g_{ij}$ are
$$\tilde{\lambda}_i := \lambda_i - \frac{1}{3} (k_{lm}g^{lm}).$$
Also note that $\sum_i (\tilde{\lambda}_i)^2 = \sigma_{lm} \sigma^{lm}$.
Then by (\ref{sigmaijij}) we get
$$\tilde{\lambda}_i = \mathcal{O}(e^{-\gamma t}).$$
Therefore with (\ref{kijgij}) we obtain
$$p_i - \frac{1}{3} = \frac{\tilde{\lambda}_i}{\frac{1}{3}(k_{lm}g^{lm})} = \mathcal{O}(e^{-\gamma t}).$$
\hfill $\Box$
%%################################################################

Before we look at the momenta of the distribution function, there is another quantity to be mentioned 
regarding expanding models, which is {\em the deceleration parameter}, say $q$. This deceleration
parameter is related to the mean curvature, as followings
$$
\partial_t (k_{ij} g^{ij}) = - (1+ q)(k_{ij} g^{ij})^2.
$$
In the inflationary models, the deceleration parameter is negative. Especially exponential expansions
lead to the value $-1$ for the deceleration parameter. Here is the statement.

%%##########################################
%% for the deceleration prarameter
%%##########################################
\begin{Proposition}
$$ q = -1 + \mathcal{O}(e^{- 2\gamma t})$$
for large $t$.
\end{Proposition}
%%##########################################
{\sc Proof.} It follows by Proposition \ref{kg+sigma}. \hfill $\Box$

%%##################################
%% characteristic
%%##################################
Now we will investigate the behaviour of the momenta of the distribution function $f$
along the characteristics where $f$ is a constant.
From the Vlasov equation (\ref{vlasov})
we define the characteristic curve $V^i (t)$ by
\begin{equation}
\frac{dV^i}{dt} = 2 k^i_j V^j - (1+g_{rs}V^r V^s)^{-1/2}\gamma^i_{mn} V^m V^n
\end{equation}
for each $V^i(\check{t}) = \check{v}^i$ given $\check{t}$.
The characteristics $V_i$, rather than $V^i$, have a bit simpler form,
which makes investigating the behaviour of the momenta easier.
So here $V_i (t)$ satisfies
\begin{equation}\label{dVi}
\frac{dV_i}{dt} = - (1+g_{rs}V^r V^s)^{-1/2}\gamma^j_{mn} V_{p} V_{q}
				  g^{pm}g^{qn}g_{ij}
\end{equation}
for each $V_i(\check{t}) = \check{v}_i$ given $\check{t}$.
For the rest of the paper, the capital $V^i$ and $V_i$ indicate that $v^i$ and $v_i$
are parameterized by the coordinate time $t$, respectively.

%%#################################################
%%           thm : Vi constant
%%#################################################
\begin{Theorem} \label{thm:V}
Consider all Bianchi types except type IX. 
Then $V_i(t)$ from (\ref{dVi}) converges to a constant along the characteristics,
as $t$ goes to infinity. I.e., $V^i(t) = e^{-2 \gamma t}\big(C+\mathcal{O}(e^{-\gamma t})\big)$ 
where $C$ is a constant, for large time $t$.  
\end{Theorem}

This theorem says that the particles in the cosmological models 
with a positive cosmological constant are expanding with a momentum whose components $v^i$ decay exponentially
for future times.

Before proving Theorem \ref{thm:V}, we require some lemmas.

%%######################
%%   g^{ij} V_i V_j
%%######################
\begin{Lemma}\label{gvv}
$$
g^{ij} V_i V_j = e^{-2 \gamma t} \big( \mathcal{V} + \mathcal{O}(e^{-\gamma t}) \big)
$$
where $\mathcal{V}$ is a constant and for large time $t$.
\end{Lemma}
%%######################
{\sc Proof.}
First note that by Propositions \ref{sigmaijt} and \ref{prop_g}, we have
$$\sigma^{ij} = \mathcal{O}(e^{-3 \gamma t}).$$
Let $\bar{\sigma}^{ij}:= e^{3 \gamma t} \sigma^{ij}$.
Then $\bar{\sigma}^{ij}$ is bounded by a constant for all $t \geq 0$.
Since $\mathcal{G}^{ij}$ in Proposition \ref{prop_g} is positive definite and time independent,
there exists a constant $C$, independent of time, such that
\begin{equation*}
\bar{\sigma}^{ij} V_i V_j \leq C \mathcal{G}^{ij} V_i V_j.
\end{equation*}
Then combining this and (\ref{kijgij}), we obtain
\begin{align}
\frac{d}{dt} (g^{ij} V_i V_j )
					 &= 2 k^{ij} V_i V_j \notag\\
				  	 &= \frac{2}{3} (k_{lm}g^{lm}) g^{ij} V_i V_j 
						   	  + 2 \sigma^{ij} V_i V_j\notag\\
					 & \leq \big(-2 \gamma + \mathcal{O}(e^{-2\gamma t})\big) g^{ij} V_i V_j 
					   		+ C e^{-3 \gamma t} \mathcal{G}^{ij} V_i V_j \notag\\
					 & \leq \big(-2 \gamma + Ce^{- \gamma t} \big)g^{ij} V_i V_j. \label{dsgvv}
\end{align}
Here to get the first equal sign, (\ref{dVi}) is used. Yet the terms involved
with (\ref{dVi}) vanish due to the antisymmetric property of $\gamma^l_{mn}$ combining with $g^{ij}$.
Now consider $V := e^{2 \gamma t} g^{ij} V_i V_j$.
Then one can see from (\ref{dsgvv}) that
\begin{equation}
\frac{dV}{dt} = \mathcal{O}(e^{- \gamma t}) V.
\end{equation}
So there exists a limit of $V$, say $\mathcal{V}$ as $t$ goes to infinity.
This gives that
\begin{equation}
g^{ij}V_i V_j = e^{-2 \gamma t} \big(\mathcal{V} + \mathcal{O}(e^{- \gamma t})\big).
\end{equation}
\hfill $\Box$

%%#####################
%%      |V_i|
%%#####################
\begin{Lemma}\label{vi}
$$
|V_i|\leq C + \mathcal{O}(e^{-\gamma t})
$$
where $C$ is a constant, and for all $i$ and for large time $t$.
\end{Lemma}
%%####################
{\sc Proof.}
Since $\mathcal{G}^{ij}$ is positive definite, there exists a constant
$C$ such that 
\begin{equation*}
|V_i|^2 \leq C \mathcal{G}^{ij} V_i V_j.
\end{equation*}
With (\ref{g^ij}), Lemma \ref{gvv} implies that
$$
\mathcal{G}^{ij} V_i V_j \leq \big( \mathcal{V} + \mathcal{O}(e^{-\gamma t}) \big).
$$
So the lemma follows. \hfill $\Box$

%%%%%%%%%%%%%%%%%%%%%%%%%%%%%%%%%%%%%%%%%%%%%%%%%%%%%%%%%%%%%%%%%%%%%%

{\sc {Proof of Theorem \ref{thm:V}}}. 
Note that in Bianchi type I, since all structure constants are zero,  also the Ricci rotation coefficients $\gamma^j_{mm}$
are zero. 
So from (\ref{dVi}) it is clear that $V_i (t) = \check{v}_i$ for all $t$, where
$\check{v}_i=V_i(\check{t})$ for a given time $\check{t}$.

More generally Lemma \ref{vi} tells us that all $V_i$ are bounded
by a constant when $t$ goes to infinity. However this allows 
oscillating behaviours. So to rule out these cases, we need to go back to
look at $\frac{dV_i}{dt}$.
Using (\ref{g_ij}) and (\ref{g^ij}), we have
\begin{equation*}
\left|\frac{dV_i}{dt} \right| \leq C e^{-2\gamma t} |V_p| |V_q|.
\end{equation*}
Here the right hand side is a summation for some $p$ and $q$.
So by Lemma \ref{vi} we obtain
\begin{equation*}
\left|\frac{dV_i}{dt}\right| \leq C e^{-2\gamma t}
\end{equation*}
where $C$ is a constant.
This leads to the conclusion that when $t$ goes to the infinity, 
$\frac{dV_i}{dt}$ goes to zero exponentially, i.e. $V_i$ goes to a constant. 
Also by (\ref{g^ij}), it is clear that $V^i(t) = e^{-2 \gamma t}\big(C+\mathcal{O}(e^{-\gamma t})\big)$. \hfill $\Box$

%%##########################################################
%% proof of geodesic completeness
%%##########################################################
Now let us prove the geodesic completeness we have postponed in the previous section.

{\sc Proof of Theorem \ref{thm.geodesic}.}
The geodesic equations for a metric of the form (\ref{metric}) imply that
along the geodesics the variables $t$, $v^i$ and $v^0$ satisfy the following
system of differential equations :
\begin{align}
\frac{dt}{d\tau} &= v^0, \label{geoeq1}\\
\frac{dv^0}{d\tau} & = k_{ij} v^i v^j, \notag\\
\frac{dv^i}{d\tau} &= 2 k^i_j v^j v^0 - \gamma^i_{mn} v^m v^n \notag
\end{align}
where $\tau$ is an affine parameter. For a particle with rest mass $m$ moving forward in time,
$v^0$ can be expressed by the remaining variables,
\begin{equation}\label{geoeq2}
 v^0 = (m^2 + g_{ij}v^i v^j)^{1/2}.
\end{equation}

The geodesic completeness is decided by looking at the relation between 
$t$ and the affine parameter $\tau$,
along any future directed causal geodesic. This relation is clear from (\ref{geoeq1}) and
(\ref{geoeq2}). I.e., it is given by
\begin{equation*} 
\frac{dt}{d\tau} = (m^2 + g_{ij} v^i v^j)^{1/2}.
\end{equation*}
To control this, we need to control $g_{ij} v^i v^j$ as a function of the coordinate time $t$.
Consider first the case of a timelike geodesic. I.e., $m >0$.
Then $V^i(t)$ satisfy
$$
\frac{dV^i}{dt} = 2 k^i_j V^j- (m^2 + g_{ij}V^i V^j)^{-1/2} \gamma^i_{mn} V^m V^n.
$$
So we can use the proof of Lemma \ref{gvv} and we get $g_{ij} V^i V^j$ is bounded.
Hence this gives that 
\begin{equation} \label{geodesic}
\frac{d\tau}{dt} = (m^2 + g^{ij} V_i V_j)^{-1/2} \geq  C
\end{equation}
where $C$ is a positive constant.
Therefore when $\tau$ is recovered by integrating (\ref{geodesic}), 
the integral of the right hand side diverges as $t$ goes to infinity.

Now consider a null geodesic, i.e., $m=0$.
Then in this case, $V^i(t)$ satisfy
$$
\frac{dV^i}{dt} = 2 k^i_j V^j - (g_{ij}V^i V^j)^{-1/2} \gamma^i_{mn} V^m V^n.
$$
So the proof of Lemma \ref{gvv} is still applied.
Again $g_{ij} V^i V^j$ is bounded and
\begin{equation*}
\frac{d\tau}{dt} = (g_{ij} V^i V^j)^{-1/2} \geq  C
\end{equation*}
where $C$ is a positive constant.
Therefore as $t$ goes to infinity so does $\tau$.
\hfill $\Box$

The rest of the paper, we analyse the energy-momentum tensor.
%%##############################################################
%% energy-momentum tensor
%%##############################################################
\begin{Proposition}
Let us consider the energy-momentum tensor in an orthonormal frame, $\{\hat{e}_i\}$. 
The energy-momentum tensor is described by
\begin{align}
\rho (t) & = \int f(t, \hat{v}) (1+|\hat{v}|^2)^{1/2} \,d\hat{v}\\
J_i (t) & = \int f(t, \hat{v}) \hat{v}_i \,d\hat{v}\\
S_{ij} (t) & = \int f(t, \hat{v}) \hat{v}_i \hat{v}_j (1+|\hat{v}|^2)^{-1/2} \,d\hat{v}
\end{align}
where $\rho:= \hat{T}_{00}$ is the energy density, $J_i := \hat{T}_{0i}$ the components of the current density
and $S_{ij} := \hat{T}_{ij}$ are the spatial components of the energy-momentum tensor.
Here the hats indicate that objects are written in the orthonormal frame.
Furthermore $\hat{v} := (\hat{v}_1, \hat{v}_2, \hat{v}_3)$ and $d \hat{v} = d\hat{v}_1 d\hat{v}_2 d\hat{v}_3$.

Then $\rho (t)$, $J_i(t)$ and $S_{ij}(t)$ tend to zero as $t$ goes to infinity.
More precisely, 
\begin{align*}
\rho (t) &= \mathcal{O}(e^{-3 \gamma t}),\\
J_i(t) &= \mathcal{O}(e^{-4 \gamma t}),\\
S_{ij}(t) &= \mathcal{O}(e^{-5 \gamma t}),
\end{align*}
for large time $t$. Furthermore
\begin{align}
\frac{J_i(t)}{\rho(t)} &= \mathcal{O}(e^{- \gamma t}),\label{ratio1}\\
\frac{S_{ij}(t)}{\rho(t)} &= \mathcal{O}(e^{- 2\gamma t}), \label{ratio2}
\end{align}
for large $t$.
\end{Proposition}

%%###############################################################
{\sc Proof.}
Combining with the fact that $f(0, v)$ has compact support on $v$,
Theorem \ref{thm:V} implies that $V_i(t)$ is {\em uniformly bounded} in large time t.
I.e., there exists a constant $C$ such that 
\begin{equation}\label{f_vi}
f(t, v)=0, \quad \text{if } |v_i| \geq C 
\end{equation}
for all large $t$.
By (\ref{g_ij}) we have
\begin{equation} \label{hat_vi}
\hat{v}_i = \mathcal{O}(e^{-\gamma t}) v_i.
\end{equation}
So using (\ref{f_vi}) and (\ref{hat_vi}) we get
\begin{equation*}
\rho (t) = \int_{|\hat{v}_i| \leq C e^{-\gamma t}} f(t, \hat{v}) (1+|\hat{v}|^2)^{1/2} \,d\hat{v}.
\end{equation*}
Note that since $f(t, \hat{v})$ is a constant along the characteristics,
\begin{equation}\label{f_f0}
|f(t, \hat{v})| \leq \|f_0\|:= \sup \{ |f(0, \hat{v})| : \forall \, \hat{v}\}.
\end{equation}
So we obtain
\begin{align*}
\rho (t) &\leq C \int_{|\hat{v}_i| \leq C e^{-\gamma t}} f(t, \hat{v}) \,d\hat{v}\\
		 &\leq C \|f_0\| e^{-3 \gamma t}.
\end{align*}
Also by (\ref{f_vi}) - (\ref{f_f0}) we have
\begin{align*}
J_i (t) &= \int_{|\hat{v}_i| \leq C e^{-\gamma t}}  f(t, \hat{v}) \hat{v}_i \,d\hat{v}\\
	 	 &\leq C e^{-\gamma t}\int_{|\hat{v}_i| \leq C e^{-\gamma t}} f(t, \hat{v}) \,d\hat{v}\\
		 &\leq C \|f_0\| e^{-4 \gamma t}.
\end{align*}
Similarly we get
\begin{align*}
S_{ij} (t) &= \int_{|\hat{v}_i| \leq C e^{-\gamma t}} f(t, \hat{v}) \hat{v}_i \hat{v}_j (1+|\hat{v}|^2)^{-1/2}  \,d\hat{v}\\
	 	 &\leq C e^{-2 \gamma t} \int_{|\hat{v}_i| \leq C e^{-\gamma t}} f(t, \hat{v}) \,d\hat{v}\\
		 &\leq C \|f_0\| e^{-5 \gamma t}.
\end{align*}
Now let us estimate the ratios $J_i/ \rho$ and $S_{ij}/\rho$. Again by (\ref{f_vi}) and (\ref{hat_vi}),
first we get
\begin{align*}
\frac{J_i(t)}{\rho(t)} 
&= \frac{\int f(t, \hat{v})  \hat{v}_i \,d\hat{v}}{\int f(t, \hat{v}) (1+|\hat{v}|^2)^{1/2} \,d\hat{v}}\\
&\leq C e^{- \gamma t} \frac{\int f(t, \hat{v}) \, d\hat{v}}{\int f(t, \hat{v}) (1+|\hat{v}|^2)^{1/2} \,d\hat{v}}\\
&\leq C e^{- \gamma t}.
\end{align*}
And similarly we obtain
\begin{align*}
\frac{S_{ij}(t)}{\rho(t)} 
&= \frac{\int f(t, \hat{v}) \hat{v}_i \hat{v}_j (1+|\hat{v}|^2)^{-1/2} \,d\hat{v}}{\int f(t, \hat{v}) (1+|\hat{v}|^2)^{1/2} \,d\hat{v}}\\
&\leq C e^{- 2\gamma t}.
\end{align*}
\hfill $\Box$
 
In this proposition since all components of the energy momentum tensor in an orthonormal frame
go to zero exponentially as $t$ goes to infinity,
one can say that in a certain sense
solutions of Einstein-Vlasov system with a positive cosmological constant
are approximated by vacuum Einstein solutions.
In a more detailed level (\ref{ratio1}) and (\ref{ratio2}) resemble the non-tilted dust-like solutions 
($J_i(t)$ and $S_{ij}(t)$ are identically zero).

\section*{Acknowlegements}
The author thanks Dr.~Alan~D.~Rendall for many valuable discussions and suggestions.

%%#########################################
%% bibliography
%%#########################################

\end{document}